\documentclass[showpacs,twocolumn,floatfix,aps,pre,superscriptaddress]{revtex4}
\usepackage{graphics}
\usepackage{tikz}
\usepackage{pgfplots}
\usepackage{amsmath}
\usepackage{natbib}
 
\newcommand{\beq}{\begin{equation}}  
\newcommand{\eeq}{\end{equation}}  
\newcommand{\bea}{\begin{eqnarray}}  
\newcommand{\eea}{\end{eqnarray}} 

\newcommand\Rey{\mbox{\textit{Re}}}  

\begin{document}

\title{Streamwise decay of localized states in channel flow}
\author{Stefan Zammert}
\affiliation{Laboratory for Aero and Hydrodynamics, TU Delft, 2628 CD Delft, The Netherlands}
\author{Bruno Eckhardt}  
\affiliation{Fachbereich Physik, Philipps-Universit\"at Marburg, D-35032 Marburg, Germany}
\affiliation{JM Burgerscentrum, TU Delft, 2628 CD Delft, The Netherlands}
\date{\today}

\begin{abstract}
Channel flow, the pressure driven flow between parallel plates, has exact coherent structures that
show various degrees of localization. For states which are localized in streamwise direction but 
extended in spanwise direction, we show that they are exponentially localized, with decay constants
that are different on the upstream and downstream side. 
We extend the analysis of Brand and Gibson [J. Fluid Mech. 750, R1 (2014)] for  
stationary states to the case of advected structures that is needed here,  
and derive expressions for the decay in terms of eigenvalues and eigenfunctions 
of certain second order differential equations. The results are in very good agreement with observations on 
exact coherent structures of different transversal wave lengths.
\end{abstract}


\maketitle

Close to onset, turbulence in subcritical shear flows is spatially and temporally intermittent.
One manifestation of the intermittency is the coexistence of laminar and turbulent regions for the
same global flow conditions that result in chaotic spatio-temporal patterns
\cite{Barkley2005,Brethouwer2012,Avila2013a,Moxey2010}.
Another is the existence of isolated localized turbulent structures which are not influenced by neighboring structures 
and can appear in the form of puffs or slugs in pipe flow \cite{Wygnanski1975} or spots in plane 
Couette \cite{Carlson1982,Lundbladh1991,Schumacher2001}  and Poiseuille flow \cite{Lemoult2013b}. 
The localized patterns and their evolution are important for an understanding
of the spatio-temporal dynamics of transitional flows \cite{Eckhardt2014a}
and in particular for the proposed link to the phenomenology of directed percolation \cite{Lemoult2016,Sano2015,Allhoff2012}, where 
a sufficiently strong decay of the interactions between structures is one of the criteria
for the selection of the appropriate universality class \cite{Hinrichsen2000}.

The relation between the transition to turbulence and exact coherent structures (ECS),
their secondary bifurcations and the formation of a dense web of dynamical links between the
different states has been studied in several flows \cite{Schmiegel1999,Gibson2009,Kawahara2012,Mellibovsky2012,Kreilos2012,Avila2013}.
For the case of plane Poiseuille flow (PPF) that is considered here, 
examples of ECS and their bifurcations include travelling waves \cite{Nagata2013a,Wall2016,Gibson2014,Zammert2015}, 
relative periodic orbits \cite{Toh2003,Zammert2014a} and streamwise \cite{Mellibovsky2015} and doubly-localized relative periodic orbits \cite{Zammert2014b}.

For an understanding of localized turbulence from its microscale, localized ECS as described in
 \cite{Schneider2010a,Schneider2010,Avila2013,Gibson2014,Brand2014,Zammert2014b} are necessary.
 The states tend to be exponentially localized in the streamwise direction, and more strongly localized
 in the spanwise direction. In very large domains, and if they are localized in both directions, 
 there are indications for algebraic  contributions as well. Specifically, 
localized ECS in plane Couette flow and plane Poiseuille flow show an exponential decay \cite{Gibson2014,Brand2014} 
which for doubly-localized states becomes algebraic for large distances \cite{Zammert2014b}. In pipe flow,
the rate of decay in streamwise direction has been shown to decrease with increasing Reynolds number \cite{Chantry2013}.

The model of Brand\&Gibson \cite{Brand2014} provides a rationale for an exponential decay and a relation
between the decay, the Reynolds number, and the spanwise modulation in the case of plane Couette flow. 
It has been tested and verified for
the streamwise velocity component in the tails of streamwise localized stationary solutions.
We here extend their model to travelling ECS in plane Poiseuille flow, and show that the 
decay rates in the upstream and downstream direction are different, 
and that this asymmetry is controlled by the advection speed of the structures.

For our studies of PPF we choose a coordinate system where the $x$-axis is oriented along the direction of the flow and the plates are 
located parallel to the $x$-$z$-plane at $y=\pm d$. The Reynolds number $Re=U_{0}d / \nu$ is defined using half the distance between the plates $d$ together with the laminar center-line velocity $U_{0}$ and the kinematic viscosity $\nu$. 
With $U_{0}$ and $d$ 
the scales for velocity and length, the laminar profile
becomes $U(y)=1-y^{2}$. We decompose the velocity field $\tilde{\textbf{u}}$ into a laminar background and a deviation,
 $\tilde{\textbf{u}}=U(y)\textbf{e}_{x}+ \textbf{u}$, and will henceforth focus on the deviations 
 $\textbf{u}=(u, v, w)$, where $u$, $v$ and $w$ are the streamwise, wall-normal and spanwise velocity components, respectively.

\begin{figure}%
	\centering
		\includegraphics[]{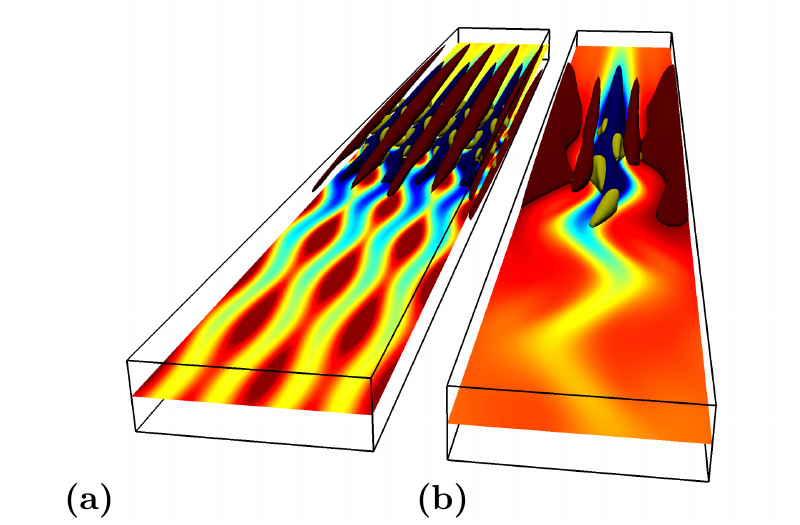}
	\caption{Visualization of the two localized ECS $PO_{Eyz}$ (a) and $PO_{E}$ (b). 
	The plots show isosurfaces of positive (red) and negative streamwise velocity and isosurfaces of the $Q$-vortex criterion (yellow)
	at $Q=0.01$ and $Q=0.001$ . 
	The entire computational domain is $402d$ long, but only a section of length $52d$ is shown in the figures.
	For both states the shown computational domain has width of $2\pi$.}
	\label{fig_Visualizations}
\end{figure}

For the numerical simulations we use the spectral code \textit{Channelflow} \citep{J.F.Gibson2012}, and
for the identification of the exact solutions 
the  Newton-method \citep{Viswanath2007} already implemented in 
channelflow.  For the linear algebra routines and the eigenvalue calculations we use the Eigen-package \citep{eigenweb}.
In all simulations, a constant mass flux is imposed. 
We use two states for the comparison between the theoretical predictions and the observed decay.
One state, henceforth referred to as $PO_{E}$, is a streamwise localized ECS relative periodic orbit obtained by  edge 
tracking \citep{Zammert2014b,Skufca2006,Schneider2007}.
The orbit is mirror-symmetric with respect to the center plane ($s_{y}$-symmetry).
By restricting the system to a subspace symmetric under wall-normal ($s_{y}$) and spanwise ($s_{z}$)
reflections, it is possible to identify a second streamwise localized relative periodic orbit, labelled $PO_{Eyz}$.
Both states have in common that they are long-wavelength instabilities \cite{Chantry2013,Melnikov2014} of 
streamwise extended travelling wave solutions:
$PO_{E}$ bifurcates subcritically from the travelling wave $TW_{E}$ \cite{Zammert2014b} and $PO_{Eyz}$ connects 
at high $\Rey$ to the travelling wave $TW_{Eyz}$ \cite{Zammert2015}.
The orbits are shown in Fig.~\ref{fig_Visualizations}.
For the present study we prescribe spanwise wavenumbers of $\gamma=1$ for $PO_{E}$  and 
of $\gamma=2$ for $PO_{Eyz}$. This allows us to study the dependence of the decay rates on the spanwise wavenumber, a 
parameter that affects the localization rate in plane Couette flow \cite{Brand2014}.
For the orbit $PO_{E}$ a numerical resolution $N_{y}\times N_{z}=49 \times 48$ with 
 $24$ modes per $2\pi$ length in streamwise direction is used.  
 For $PO_{Eyz}$ we use $N_{y}\times N_{z}=49 \times 32$ and $32$ modes per $2\pi$ in streamwise direction.

\begin{figure}
	\centering
		\includegraphics[]{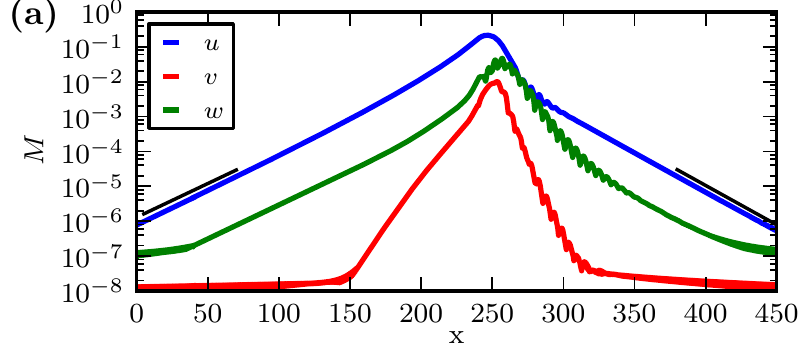}
		\includegraphics[]{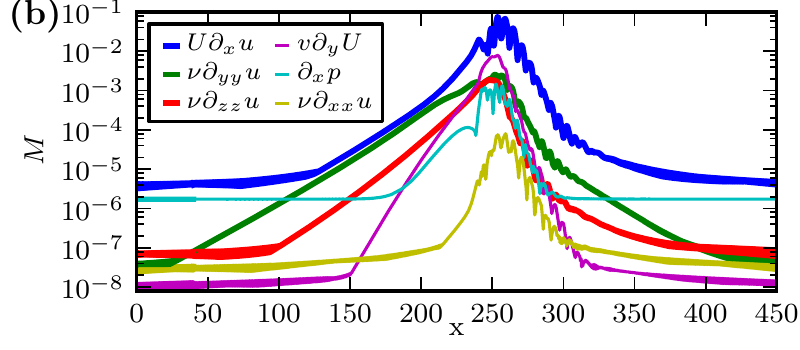}
	\caption{Localization properties of the state $PO_E$ at $\Rey=1100$. 
	(a) 
	maximum $M$ of the absolute values of the velocity components is shown versus streamwise coordinate $x$.  
	Black lines show exponential decays with decays constants obtained from the model for the upstream and 
	downstream tails.
	(b) the maximum norm of the terms of the linearized equations which contribute to the 
	change of the streamwise component are shown.}
	\label{fig_LinfTerms}
\end{figure}

We quantify the localization properties using the maximum values 
of the velocity components as a function of streamwise
coordinate $x$, with the maximum taken over spanwise and wall-normal positions,
as shown in Fig. \ref{fig_LinfTerms}(a) for the state $PO_{E}$.
As in previous studies \cite{Brand2014,Zammert2014b}, all three velocity components decay exponentially, but since
there is no symmetry in the flow direction, the decay rates in the upstream and downstream direction differ in general.
One also notes that the decay of the streamwise and spanwise components is similar, but different from the one in the normal
velocity component.

The key to the explanation of the decay rates is the observation that away from the center of the ECS, the velocity components
are small and can be described with a linearization of the Navier-Stokes equation around the laminar profile. The terms that
remain for the perturbation $\textbf{u}$ are
\begin{equation}
\partial_{t} \textbf{u} + U(y) \partial_{x} \textbf{u} + v \partial_{y} U(y) \textbf{e}_{x}= -\nabla p + \frac{1}{Re} \Delta \textbf{u}
\end{equation}
Here, $U(y)=1-y^2$ is the laminar velocity profile and $p$
the pressure fluctuations in the perturbation. In order to separate the contributions of each term to the
equation,  we show in Fig.~\ref{fig_LinfTerms}(b) the maximum over spanwise and wall-normal positions of the different terms as a function of streamwise positions $x$.
The data shows that in the tails we can neglect the coupling to the wall-normal velocity and to the base flow 
($ v \partial_{y} U$), as well as the second order derivative $\nu \partial_{xx}u$.
We also neglect the pressure fluctuations since the plateau value which can be seen in Fig.~\ref{fig_LinfTerms}(b) is related to 
the periodic boundary conditions and decreases with increasing domain length.

Without these terms, the equation for the streamwise component reduces to
\begin{equation}
\partial_{t}u+U(y)\partial_{x}u=\textit{Re}^{-1}(\partial_{yy}u +  \partial_{zz}u) \label{eq1}\,.
\end{equation}
This equation is the plane Poiseuille flow equivalent of
the  equation obtained by Brand\&Gibson \cite{Brand2014} for the tails of localized states in plane Couette flow.
It shows that the form of the equation is primarily a consequence of the Cartesian geometry and of the linearization
around the laminar profile.

The 
reflection symmetry on the midplane in plane Couette flow allows for the existence of stationary solutions, 
of the type analyzed in \cite{Brand2014}. In the absence of this symmetry, all ECS for PPF are advected in streamwise direction. The travelling waves in plane Couette flow described in \cite{Nagata1997} also lack this symmetry
and the analysis we describe here should also apply to localized versions of these travelling waves.
Therefore, the ansatz for the flow field in the tails has to include the advection, and we take 
a spanwise periodic and streamwise travelling field of the form  
\begin{equation}
u(x, y,z,t) =u(y) \exp\left[i\gamma z-\mu(x-ct)/Re\right].
\label{ansatz}
\end{equation}
In this ansatz $\gamma$ is the spanwise wavenumber, $c$ is the wave speed and $\mu$ is
the decay rate that we want to calculate.

\begin{figure}
\includegraphics[]{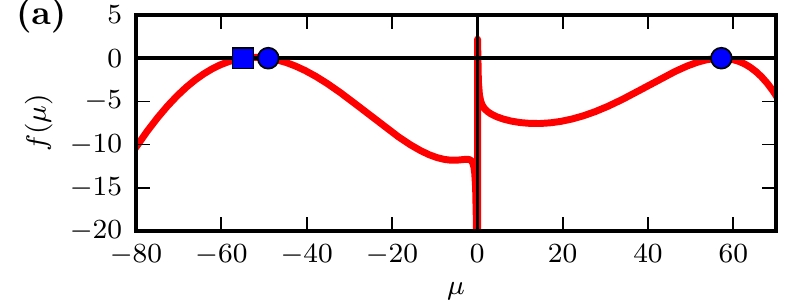} 
\includegraphics[]{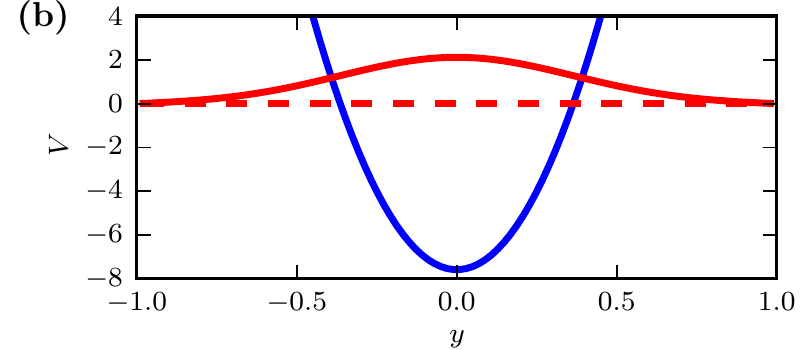} 
\includegraphics[]{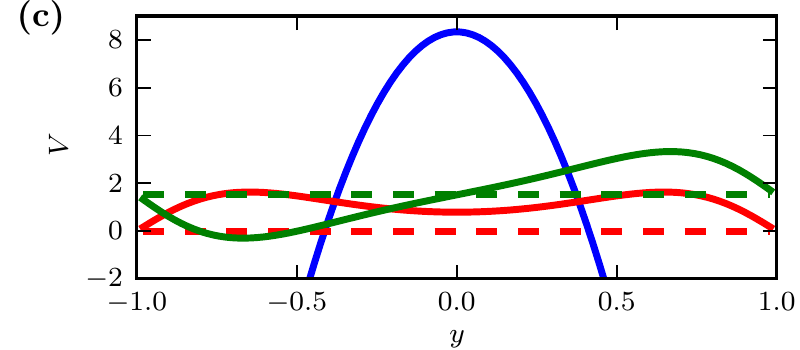} 
\caption{Determination of the decay exponent $\mu$.
Panel (a) shows $f(\mu)$ vs. $\mu$ for $\gamma=1.0$ and $c=0.85$. 
The zeros are marked by blue symbols.
Circles and squares are used if the corresponding profile is symmetric and asymmetric, respectively.
At $\mu=413$ (not shown) another zero with an asymmetric profile exists.
Panel (b)  shows the potential $V(y)$ for the downstream decay, 
for $\gamma=1.0$, $c=0.85$ and $\mu=57.26$ (blue),
where a symmetric eigenfunction with eigenvalue $0$ exists (red).
Panel (c) shows the potential for the parameters $\gamma=1.0$, $c=0.85$ and $\mu=-48.97$ (blue)
on the upstream side. Here,
$0$ is an eigenvalue with a symmetric eigenfunction (red), while the 
 asymmetric eigenfunction (green) corresponds to an eigenvalue of $1.51$. 
The dashed lines in (b) and (c)  indicate the positions of the eigenvalues for the eigenfunctions of the same color.
\label{fig_fmu}}
\end{figure}

Substituting the ansatz and the PPF baseflow in Eq.~\ref{eq1} then gives
\begin{equation}
u'' - (\gamma^{2} + \mu (c+y^{2} -1))u=0
\label{u_eq}
\end{equation}
and we have to seek solutions that satisfy the friction boundary conditions $u(\pm 1)=0$.
A general solution of this second order differential equation can be given in terms of 
parabolic cylinder functions $D_\nu(z)$ \cite{Jeffrey},
\begin{equation}
u(y)=a \cdot  D_{\nu_1}(\sqrt{2} \sqrt[4]{\mu} y) 
+ 
b \cdot D_{\nu_2}(i\sqrt{2} \sqrt[4]{\mu} y) .
\end{equation}
with indices $\nu_1=(-\gamma^{2}-c\mu+\mu-\sqrt{\mu})/(2\sqrt{\mu})$ and 
$\nu_2=(\gamma^{2}+(c-1)\mu-\sqrt{\mu})/(2\sqrt{\mu})$.
The boundary conditions $u(\pm1)=0$ provide a set of linear equations that only has non-trivial 
solution if the discriminant
\begin{multline}
f(\mu)=D_{\nu_1} (-\sqrt{2} \sqrt[4]{\mu})\cdot D_{\nu_2}(i\sqrt{2} \sqrt[4]{\mu} )  \\
-D_{\nu_1}(\sqrt{2} \sqrt[4]{\mu})\cdot D_{\nu_2}(-i\sqrt{2} \sqrt[4]{\mu} ).
\label{EquationForMu}
\end{multline}
vanishes. This function is readily evaluated using suitable numerical software, 
and one can compute or read off the zeros 
from curves like the ones shown in Fig.~\ref{fig_fmu}(a).
The specific case in Fig.~\ref{fig_fmu}(a) uses the parameters $c=0.85$ 
(the average wavespeed of $PO_{E}$ at $\Rey=1100$) and $\gamma=1.0$, 
and has zeros at $\mu = -54.9$, $-48.97$ and $57.26$.

However, the properties of the eigenvalues and the
eigenfunctions become more transparent when considering the relation between (\ref{u_eq}) and 
an eigenvalue problem for a one-dimensional Schr\"odinger equation with a potential $V(y)$,
\begin{equation}
-u'' + V(y) u = E u.
\label{Seq}
\end{equation}
Eq. \ref{u_eq} is then reinterpreted as an eigenvalue problem for (\ref{Seq}) with potential 
$V(y) = \gamma^2+\mu(c-1) + \mu y^2$, boundary conditions $u(\pm 1)=0$, and 
eigenvalue $E=0$. Since $\gamma$ and $c$ are prescribed, the task is to determine $\mu$ such
that the eigenvalue problem has a consistent solution. 
Examples of the potential and its eigenvalues are shown in Figs. \ref{fig_fmu}(b) and (c).
This form of the eigenvalue
problem allows some qualitative conclusions and approximations to the decay rates
and the shape of the profiles. 
As the type of the potential changes when $\mu$ changes sign, we consider the downstream
and upstream situations separately.

\begin{figure}
	\centering
	\includegraphics[]{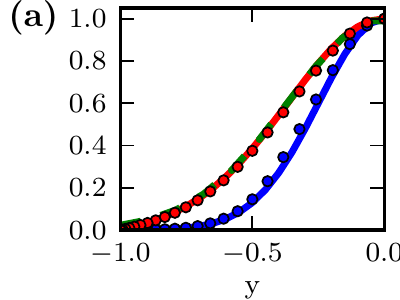}\includegraphics[]{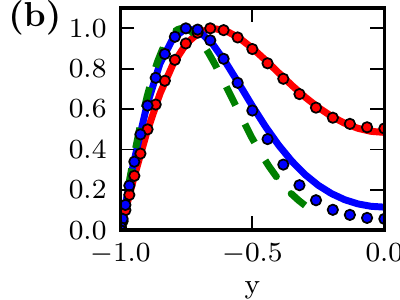}
\caption{Streamwise velocity profiles on the downstream (a) and upstream (b) side for the states 
$PO_{E}$ (red) and $PO_{Eyz}$ (blue) at $\Rey=1100$ and $\Rey=1800$, respectively.
Circles represent DNS data and solid lines are used for the model profiles.
In addition, the dashed green lines show a gaussian $\exp{(-\kappa y^2 /2)}$ in (a), and an 
Airy function in (b), as approximate eigenfunctions for the corresponding eigenvalue problems for $PO_{E}$.
Since the profiles are symmetric to the centerline, only one half of the channel width is shown.}
\label{fig_DNSvsModelProfiles}
\end{figure}

On the downstream side, $\mu=\kappa^2$ is positive. Then the potential
$V(y)$ is a harmonic oscillator with a minimum at $y=0$ of 
depth $V_{min}=\gamma^2+(c-1)\kappa^2$
and curvature $V''=2\kappa^2$. The minimum of the potential has to be 
negative, so that $c<1$ and $\kappa^2>\gamma^2/(1-c)$.
The region where the potential is larger than the eigenvalue, $V(y)>0$, is the 
classically forbidden region where the solutions usually fall off quickly. So if the boundary
points $y=\pm1$ are sufficiently far inside the forbidden region they may be moved out
to $\pm \infty$ with little effect on the eigenvalues (and the eigenfunctions).
For the case shown in Figs. \ref{fig_fmu}(b) the forbidden region starts near $|x|\approx0.4$
and the approximation to consider the boundary conditions $u(\pm\infty)=0$ is justified.
The ground state for such a harmonic potential has a Gaussian form,
$u\propto \exp(-\kappa y^2/2)$ and an eigenvalue that lies at a value $\kappa$ above the minimum,
so that the equation for $\kappa$ becomes $V_{min}+\kappa=0$ or
\begin{equation}
\gamma^2-(1-c)\kappa^2+\kappa = 0.
\end{equation}
The relevant solution of this quadratic equation is
\begin{equation}
\kappa_0=\frac{1}{2(1-c)}+\sqrt{\frac{\gamma^2}{1-c}+\frac{1}{4(1-c)^2}}.
\end{equation}
For the specific case shown in Fig. \ref{fig_fmu}, this results in the approximate values 
$\mu_0\approx 57$, in 
excellent agreement with the value $\mu=57.26$ from the exact expression. 
The calculation also shows that the
eigenfunction can be well approximated by a Gaussian controlled by $\kappa\approx 7.55$, as shown
by the dashed green line in  Fig. \ref{fig_DNSvsModelProfiles}(a).

On the upstream side, $\mu$ is negative,
and the potential is an inverted parabola with a maximum
in the middle, $V_{max}=\gamma^2+(1-c) |\mu|$. 
The potential assumes its minima at the walls, $V_{min}=\gamma^2-|\mu| c$,
and since these minima have to be negative, we have $|\mu|>\gamma^2/c$. 
The potential starts with a linear slope away from the wall, so that the 
approximate potential has the form $V\approx V_{min} + 2 |\mu| \tilde y$, where $\tilde y=y+1$ is the
distance from the wall at $y=-1$. Proceeding as in the previous case
one notes that the corresponding reference eigenvalue problem is that for 
eigenstates in a linear potential, with boundary conditions $u(0)=u(\infty)=0$.
This problem has Airy functions as its basic solutions, and a first eigenvalue at $E_1(2|\mu|)^{2/3}$ 
above the potential minimum,
where $E_1=2.338$ is the numerical value in a potential with slope $1$, \cite{Flugge1999}.
Accordingly, the determining equation for $\mu$ becomes
\begin{equation}
\gamma^2- |\mu| c + E_1 (2 |\mu|)^{2/3}=0,
\end{equation}
with the approximate solution $|\mu|\approx (4.676/c)^3(1+0.029 c^2\gamma^2)$.
For $TW_{E}$ at $Re=1100$ the decay rate from the approximate 
model becomes $\mu=-86$, which differs considerably from the 
value obtained from the exact equation.
The reason for this discrepancy is apparent from the form of the potential:
the classically forbidden region lies in the center, but the value of the potential
is not very high, so that the profile still maintains a significant value near $0$ before
it enters the other boundary layer. The approximate expression for the eigenvalues
shows, however, that this situation will change when the structures are closer to the wall
where the wave speed $c$ becomes smaller and the damping $|\mu|$ stronger. 
Similarly, also larger values of $\gamma$ will give stronger localization near the wall
and along the streamwise direction.


\begin{figure}
	\centering
	\includegraphics[]{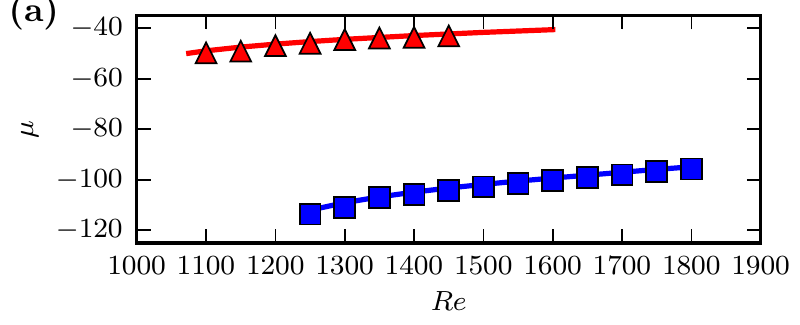} \\
	\includegraphics[]{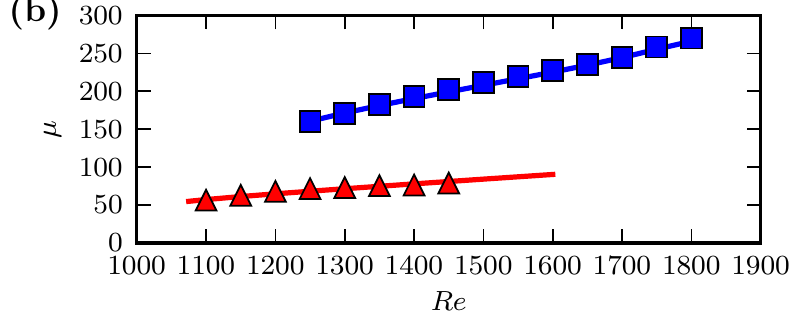} \\
	\includegraphics[]{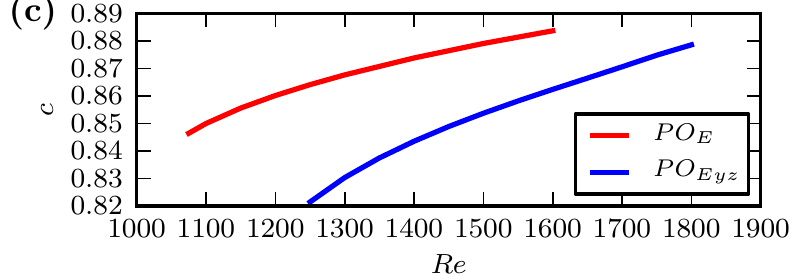}
\caption{Decay constants for exact coherent states as a function of Reynolds number.
(a) Upstream side for $PO_E$ (red) and $PO_{Eyz}$ (blue). The model predictions, which are obtained from roots 
of Eq. \ref{EquationForMu}, are  shown as continuous lines,
the decay rates fitted to the DNS data as triangles and squares, respectively. 
(b) Same as (a) for the downstream side.
(c) Average wavespeed for both states. The variation in wavespeed with $\Rey$ 
is responsible for the variation in decay constants.}
\label{fig_DNSvsModelTailHead}
\end{figure}

In order to compare the predictions for the decay from the model (\ref{eq1}-\ref{EquationForMu}) to the DNS data for the 
two states $PO_{E}$ and $PO_{Eyz}$ 
we have to extract decay
rates from the DNS data.
For the decay constants we fit an exponential  to the spanwise wall-normal maximum of the streamwise  
velocity component on the downstream and 
upstream sides of the structures. To avoid congestion by transients,
we only use data that are at least $100d$ (upstream side) and $50d$ (downstream side) 
from the center of the structures. The domains used in the simulations are therefore 
necessarily very long.
For the orbit $PO_{E}$ the decay constants were calculated in a domain with a streamwise length 
$L_{x}$ of $565h$.
For shorter domains  
decay rates that are smaller in absolute value, corresponding to a slower decay, are observed. 
The rates for the tails of the two structures and for different Reynolds numbers 
are shown in Figs.~\ref{fig_DNSvsModelTailHead}(a) and (b) using solid symbols.

The spanwise wavenumber $\gamma$ that enters the model represents the wavenumbers
in the structures that dominate the tails, and does not need to coincide with the wavenumber of the
domain. We therefore compared the observed decay with decays computed from (\ref{EquationForMu}) for different
values of $\gamma$. 
For $PO_{E}$ we find that the best fit is achieved for $\gamma=1$, the wavenumber of the
domain.
For the mirror symmetric state $PO_{Eyz}$ the best agreement with the numerical data 
is achieved for $\gamma=4$, twice the value of the domain size. 
This decrease in wavelength in the tail is induced by the 
spanwise mirror symmetry of the structure and reflected in the presence of four streaks in Fig.~\ref{fig_Visualizations}(a)

The ansatz for the velocity field in (\ref{ansatz}) describes the decay in downstream direction
as $\exp(-\mu x /\Rey)$, and suggests that both the upstream and downstream length of the
structures should increase proportional to $\Rey$. 
However, the eigenvalue problem (\ref{u_eq}) for $\mu$ contains an implicit 
$\Rey$ dependence due to the variation of the wave speed $c(Re)$. 
As shown in Fig.~\ref{fig_DNSvsModelTailHead}(c),
the wave speed $c$, determined as the average wave speed of an ECS 
calculated by dividing the displacement over one period by the length of the period, increases 
with Reynolds number. 
It is only once this dependence is taken into account
that the localization rates determined from the DNS and from 
the model give the very good agreement 
shown in Figs.~\ref{fig_DNSvsModelTailHead}(a) and (b).
The model data are also shown in 
Fig.~\ref{fig_LinfTerms}(a) 
where a black line indicates the theoretical prediction for the 
streamwise component (blue).
\begin{figure}
	\centering
	\includegraphics[]{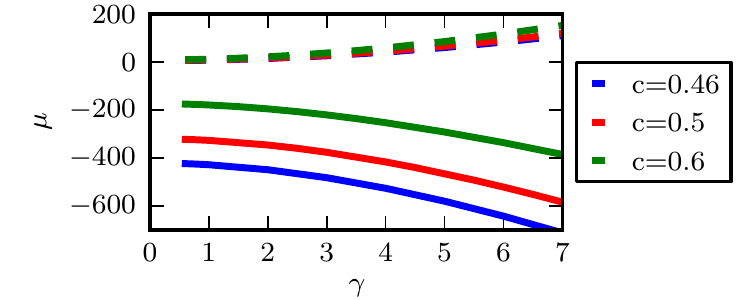}
\caption{Dependence of the decay constants $\mu$ on the spanwise wavenumber $\gamma$ for different wavespeeds $c$
for the symmetric profile on the upstream (solid) and downstream side (dashed).
}
\label{fig_mu_vs_ga}
\end{figure}
The effects of the Reynolds number dependence on the
localization length of the structures are different on the upstream and downstream side.
On the upstream side, Fig.~\ref{fig_DNSvsModelTailHead}(a) shows that $\mu$ becomes less
negative with increasing $\Rey$. Accordingly, the localization length, defined as $\Rey / |\mu|$, 
increases with $\Rey$ and the structures become stretched out along the wall. 
On the downstream side, Fig.~\ref{fig_DNSvsModelTailHead}(b)
shows that $\mu(\Rey)$ increases, roughly proportional to $\Rey$, so that the 
localization length $\Rey / |\mu|$ varies only little. 
This is in agreement with observations in \cite{Zammert2014b}.
The localization properties of modes with different spanwise wavenumbers $\gamma$ are shown in Fig.~\ref{fig_mu_vs_ga}. 
The increasing absolute value $|\mu|$ on the upstream side indicates more strongly localized
structures, whereas there are little changes on the downstream side. 
The wave speeds used in Fig.~\ref{fig_mu_vs_ga} are near $1/2$. In the limit $c\rightarrow 1$, which
may be obtained for structures that are localized in the center, and in the limit $c\rightarrow 0$ for structures
that are localized very close to the walls, the eigenvalues diverge and the structures should become strongly
localized (in the downstream and upstream direction, respectively). However, we do not have examples
of flow structures on which to verify this prediction.

The present extension of the Brand\&Gibson \cite{Brand2014} analysis to advective structures in PPF 
shows how the decay and the profile of the structures on the downstream and upstream side are 
related to the wave speed and spanwise wavenumber. Similar analyses should be possible for other
advected structures, such as localized version of travelling waves in plane Couette flow \cite{Nagata1997} 
or localized structures in the asymptotic suction boundary layer \cite{Khapko2016a}.

This work was supported in part by the DFG within FOR 1182 and by 
Stichting voor Fundamenteel Onderzoek der Materie (FOM) within the
program ``Towards ultimate turbulence''.


\begin{thebibliography}{42}
\expandafter\ifx\csname natexlab\endcsname\relax\def\natexlab#1{#1}\fi
\expandafter\ifx\csname bibnamefont\endcsname\relax
  \def\bibnamefont#1{#1}\fi
\expandafter\ifx\csname bibfnamefont\endcsname\relax
  \def\bibfnamefont#1{#1}\fi
\expandafter\ifx\csname citenamefont\endcsname\relax
  \def\citenamefont#1{#1}\fi
\expandafter\ifx\csname url\endcsname\relax
  \def\url#1{\texttt{#1}}\fi
\expandafter\ifx\csname urlprefix\endcsname\relax\def\urlprefix{URL }\fi
\providecommand{\bibinfo}[2]{#2}
\providecommand{\eprint}[2][]{\url{#2}}

\bibitem[{\citenamefont{Barkley and Tuckerman}(2005)}]{Barkley2005}
\bibinfo{author}{\bibfnamefont{D.}~\bibnamefont{Barkley}} \bibnamefont{and}
  \bibinfo{author}{\bibfnamefont{L.}~\bibnamefont{Tuckerman}},
  \bibinfo{journal}{Phys. Rev. Lett.} \textbf{\bibinfo{volume}{94}},
  \bibinfo{pages}{014502} (\bibinfo{year}{2005}).

\bibitem[{\citenamefont{Brethouwer et~al.}(2012)\citenamefont{Brethouwer,
  Duguet, and Schlatter}}]{Brethouwer2012}
\bibinfo{author}{\bibfnamefont{G.}~\bibnamefont{Brethouwer}},
  \bibinfo{author}{\bibfnamefont{Y.}~\bibnamefont{Duguet}}, \bibnamefont{and}
  \bibinfo{author}{\bibfnamefont{P.}~\bibnamefont{Schlatter}},
  \bibinfo{journal}{J. Fluid Mech.} \textbf{\bibinfo{volume}{704}},
  \bibinfo{pages}{137} (\bibinfo{year}{2012}).

\bibitem[{\citenamefont{Avila and Hof}(2013)}]{Avila2013a}
\bibinfo{author}{\bibfnamefont{M.}~\bibnamefont{Avila}} \bibnamefont{and}
  \bibinfo{author}{\bibfnamefont{B.}~\bibnamefont{Hof}},
  \bibinfo{journal}{Phys. Rev. E} \textbf{\bibinfo{volume}{87}},
  \bibinfo{pages}{063012} (\bibinfo{year}{2013}).

\bibitem[{\citenamefont{Moxey and Barkley}(2010)}]{Moxey2010}
\bibinfo{author}{\bibfnamefont{D.}~\bibnamefont{Moxey}} \bibnamefont{and}
  \bibinfo{author}{\bibfnamefont{D.}~\bibnamefont{Barkley}},
  \bibinfo{journal}{Proc. Natl. Acad. Sci. U. S. A.}
  \textbf{\bibinfo{volume}{107}}, \bibinfo{pages}{8091} (\bibinfo{year}{2010}).

\bibitem[{\citenamefont{Wygnanski et~al.}(1975)\citenamefont{Wygnanski,
  Sokolov, and Friedman}}]{Wygnanski1975}
\bibinfo{author}{\bibfnamefont{I.}~\bibnamefont{Wygnanski}},
  \bibinfo{author}{\bibfnamefont{M.}~\bibnamefont{Sokolov}}, \bibnamefont{and}
  \bibinfo{author}{\bibfnamefont{D.}~\bibnamefont{Friedman}},
  \bibinfo{journal}{J. Fluid Mech.} \textbf{\bibinfo{volume}{69}},
  \bibinfo{pages}{283} (\bibinfo{year}{1975}).

\bibitem[{\citenamefont{Carlson et~al.}(1982)\citenamefont{Carlson, Widnall,
  and Peeters}}]{Carlson1982}
\bibinfo{author}{\bibfnamefont{D.~R.} \bibnamefont{Carlson}},
  \bibinfo{author}{\bibfnamefont{S.~E.} \bibnamefont{Widnall}},
  \bibnamefont{and} \bibinfo{author}{\bibfnamefont{M.~F.}
  \bibnamefont{Peeters}}, \bibinfo{journal}{J. Fluid Mech.}
  \textbf{\bibinfo{volume}{121}}, \bibinfo{pages}{487} (\bibinfo{year}{1982}).

\bibitem[{\citenamefont{Lundbladh and Johansson}(1991)}]{Lundbladh1991}
\bibinfo{author}{\bibfnamefont{A.}~\bibnamefont{Lundbladh}} \bibnamefont{and}
  \bibinfo{author}{\bibfnamefont{A.~V.} \bibnamefont{Johansson}},
  \bibinfo{journal}{J. Fluid Mech.} \textbf{\bibinfo{volume}{229}},
  \bibinfo{pages}{499} (\bibinfo{year}{1991}).

\bibitem[{\citenamefont{Schumacher and Eckhardt}(2001)}]{Schumacher2001}
\bibinfo{author}{\bibfnamefont{J.}~\bibnamefont{Schumacher}} \bibnamefont{and}
  \bibinfo{author}{\bibfnamefont{B.}~\bibnamefont{Eckhardt}},
  \bibinfo{journal}{Phys. Rev. E} \textbf{\bibinfo{volume}{63}},
  \bibinfo{pages}{046307} (\bibinfo{year}{2001}).

\bibitem[{\citenamefont{Lemoult et~al.}(2014)\citenamefont{Lemoult, Gumowski,
  Aider, and Wesfreid}}]{Lemoult2013b}
\bibinfo{author}{\bibfnamefont{G.}~\bibnamefont{Lemoult}},
  \bibinfo{author}{\bibfnamefont{K.}~\bibnamefont{Gumowski}},
  \bibinfo{author}{\bibfnamefont{J.-L.} \bibnamefont{Aider}}, \bibnamefont{and}
  \bibinfo{author}{\bibfnamefont{J.~E.} \bibnamefont{Wesfreid}},
  \bibinfo{journal}{Eur. Phys. J. E} \textbf{\bibinfo{volume}{37}}
  (\bibinfo{year}{2014}).

\bibitem[{\citenamefont{Eckhardt}(2014)}]{Eckhardt2014a}
\bibinfo{author}{\bibfnamefont{B.}~\bibnamefont{Eckhardt}},
  \bibinfo{journal}{J. Fluid Mech.} \textbf{\bibinfo{volume}{758}},
  \bibinfo{pages}{1} (\bibinfo{year}{2014}).

\bibitem[{\citenamefont{Lemoult et~al.}(2016)\citenamefont{Lemoult, Shi, Avila,
  Jalikop, Avila, and Hof}}]{Lemoult2016}
\bibinfo{author}{\bibfnamefont{G.}~\bibnamefont{Lemoult}},
  \bibinfo{author}{\bibfnamefont{L.}~\bibnamefont{Shi}},
  \bibinfo{author}{\bibfnamefont{K.}~\bibnamefont{Avila}},
  \bibinfo{author}{\bibfnamefont{S.~V.} \bibnamefont{Jalikop}},
  \bibinfo{author}{\bibfnamefont{M.}~\bibnamefont{Avila}}, \bibnamefont{and}
  \bibinfo{author}{\bibfnamefont{B.}~\bibnamefont{Hof}}, \bibinfo{journal}{Nat.
  Phys.} \textbf{\bibinfo{volume}{12}}, \bibinfo{pages}{254}
  (\bibinfo{year}{2016}).

\bibitem[{\citenamefont{Sano and Tamai}(2015)}]{Sano2015}
\bibinfo{author}{\bibfnamefont{M.}~\bibnamefont{Sano}} \bibnamefont{and}
  \bibinfo{author}{\bibfnamefont{K.}~\bibnamefont{Tamai}},
  \bibinfo{journal}{Nat. Phys.} \textbf{\bibinfo{volume}{12}},
  \bibinfo{pages}{249} (\bibinfo{year}{2015}).

\bibitem[{\citenamefont{Allhoff and Eckhardt}(2012)}]{Allhoff2012}
\bibinfo{author}{\bibfnamefont{K.~T.} \bibnamefont{Allhoff}} \bibnamefont{and}
  \bibinfo{author}{\bibfnamefont{B.}~\bibnamefont{Eckhardt}},
  \bibinfo{journal}{Fluid Dyn. Res.} \textbf{\bibinfo{volume}{44}},
  \bibinfo{pages}{031201} (\bibinfo{year}{2012}).

\bibitem[{\citenamefont{Hinrichsen}(2000)}]{Hinrichsen2000}
\bibinfo{author}{\bibfnamefont{H.}~\bibnamefont{Hinrichsen}},
  \bibinfo{journal}{Adv. Phys.} \textbf{\bibinfo{volume}{49}},
  \bibinfo{pages}{815} (\bibinfo{year}{2000}).

\bibitem[{\citenamefont{Schmiegel}(1999)}]{Schmiegel1999}
\bibinfo{author}{\bibfnamefont{A.}~\bibnamefont{Schmiegel}}, \bibinfo{type}{PhD
  thesis}, \bibinfo{school}{Philipps-Universitit{\"{a}}t Marburg}
  (\bibinfo{year}{1999}),
  \urlprefix\url{http://archiv.ub.uni-marburg.de/diss/z2000/0062/pdf/z2000-0062.pdf}.

\bibitem[{\citenamefont{Gibson et~al.}(2009)\citenamefont{Gibson, Halcrow, and
  Cvitanovi{\'{c}}}}]{Gibson2009}
\bibinfo{author}{\bibfnamefont{J.~F.} \bibnamefont{Gibson}},
  \bibinfo{author}{\bibfnamefont{J.}~\bibnamefont{Halcrow}}, \bibnamefont{and}
  \bibinfo{author}{\bibfnamefont{P.}~\bibnamefont{Cvitanovi{\'{c}}}},
  \bibinfo{journal}{J. Fluid Mech.} \textbf{\bibinfo{volume}{638}},
  \bibinfo{pages}{243} (\bibinfo{year}{2009}).

\bibitem[{\citenamefont{Kawahara et~al.}(2012)\citenamefont{Kawahara, Uhlmann,
  and van Veen}}]{Kawahara2012}
\bibinfo{author}{\bibfnamefont{G.}~\bibnamefont{Kawahara}},
  \bibinfo{author}{\bibfnamefont{M.}~\bibnamefont{Uhlmann}}, \bibnamefont{and}
  \bibinfo{author}{\bibfnamefont{L.}~\bibnamefont{van Veen}},
  \bibinfo{journal}{Annu. Rev. Fluid Mech.} \textbf{\bibinfo{volume}{44}},
  \bibinfo{pages}{203} (\bibinfo{year}{2012}).

\bibitem[{\citenamefont{Mellibovsky and Eckhardt}(2012)}]{Mellibovsky2012}
\bibinfo{author}{\bibfnamefont{F.}~\bibnamefont{Mellibovsky}} \bibnamefont{and}
  \bibinfo{author}{\bibfnamefont{B.}~\bibnamefont{Eckhardt}},
  \bibinfo{journal}{J. Fluid Mech.} \textbf{\bibinfo{volume}{709}},
  \bibinfo{pages}{149} (\bibinfo{year}{2012}).

\bibitem[{\citenamefont{Kreilos and Eckhardt}(2012)}]{Kreilos2012}
\bibinfo{author}{\bibfnamefont{T.}~\bibnamefont{Kreilos}} \bibnamefont{and}
  \bibinfo{author}{\bibfnamefont{B.}~\bibnamefont{Eckhardt}},
  \bibinfo{journal}{Chaos} \textbf{\bibinfo{volume}{22}},
  \bibinfo{pages}{047505} (\bibinfo{year}{2012}).

\bibitem[{\citenamefont{Avila et~al.}(2013)\citenamefont{Avila, Mellibovsky,
  Roland, and Hof}}]{Avila2013}
\bibinfo{author}{\bibfnamefont{M.}~\bibnamefont{Avila}},
  \bibinfo{author}{\bibfnamefont{F.}~\bibnamefont{Mellibovsky}},
  \bibinfo{author}{\bibfnamefont{N.}~\bibnamefont{Roland}}, \bibnamefont{and}
  \bibinfo{author}{\bibfnamefont{B.}~\bibnamefont{Hof}},
  \bibinfo{journal}{Phys. Rev. Lett.} \textbf{\bibinfo{volume}{110}},
  \bibinfo{pages}{224502} (\bibinfo{year}{2013}).

\bibitem[{\citenamefont{Nagata and Deguchi}(2013)}]{Nagata2013a}
\bibinfo{author}{\bibfnamefont{M.}~\bibnamefont{Nagata}} \bibnamefont{and}
  \bibinfo{author}{\bibfnamefont{K.}~\bibnamefont{Deguchi}},
  \bibinfo{journal}{J. Fluid Mech} \textbf{\bibinfo{volume}{735}},
  \bibinfo{pages}{R4} (\bibinfo{year}{2013}).

\bibitem[{\citenamefont{Wall and Nagata}(2016)}]{Wall2016}
\bibinfo{author}{\bibfnamefont{D.~P.} \bibnamefont{Wall}} \bibnamefont{and}
  \bibinfo{author}{\bibfnamefont{M.}~\bibnamefont{Nagata}},
  \bibinfo{journal}{J. Fluid Mech.} \textbf{\bibinfo{volume}{788}},
  \bibinfo{pages}{444} (\bibinfo{year}{2016}).

\bibitem[{\citenamefont{Gibson and Brand}(2014)}]{Gibson2014}
\bibinfo{author}{\bibfnamefont{J.~F.} \bibnamefont{Gibson}} \bibnamefont{and}
  \bibinfo{author}{\bibfnamefont{E.}~\bibnamefont{Brand}}, \bibinfo{journal}{J.
  Fluid Mech.} \textbf{\bibinfo{volume}{745}}, \bibinfo{pages}{25}
  (\bibinfo{year}{2014}).

\bibitem[{\citenamefont{Zammert and Eckhardt}(2015)}]{Zammert2015}
\bibinfo{author}{\bibfnamefont{S.}~\bibnamefont{Zammert}} \bibnamefont{and}
  \bibinfo{author}{\bibfnamefont{B.}~\bibnamefont{Eckhardt}},
  \bibinfo{journal}{Phys. Rev. E} \textbf{\bibinfo{volume}{91}},
  \bibinfo{pages}{041003(R)} (\bibinfo{year}{2015}).

\bibitem[{\citenamefont{Toh and Itano}(2003)}]{Toh2003}
\bibinfo{author}{\bibfnamefont{S.}~\bibnamefont{Toh}} \bibnamefont{and}
  \bibinfo{author}{\bibfnamefont{T.}~\bibnamefont{Itano}}, \bibinfo{journal}{J.
  Fluid Mech.} \textbf{\bibinfo{volume}{481}}, \bibinfo{pages}{67}
  (\bibinfo{year}{2003}).

\bibitem[{\citenamefont{Zammert and
  Eckhardt}(2014{\natexlab{a}})}]{Zammert2014a}
\bibinfo{author}{\bibfnamefont{S.}~\bibnamefont{Zammert}} \bibnamefont{and}
  \bibinfo{author}{\bibfnamefont{B.}~\bibnamefont{Eckhardt}},
  \bibinfo{journal}{Fluid Dyn. Res.} \textbf{\bibinfo{volume}{46}},
  \bibinfo{pages}{041419} (\bibinfo{year}{2014}{\natexlab{a}}).

\bibitem[{\citenamefont{Mellibovsky and Meseguer}(2015)}]{Mellibovsky2015}
\bibinfo{author}{\bibfnamefont{F.}~\bibnamefont{Mellibovsky}} \bibnamefont{and}
  \bibinfo{author}{\bibfnamefont{A.}~\bibnamefont{Meseguer}},
  \bibinfo{journal}{J. Fluid Mech.} \textbf{\bibinfo{volume}{779}},
  \bibinfo{pages}{R1} (\bibinfo{year}{2015}).

\bibitem[{\citenamefont{Zammert and
  Eckhardt}(2014{\natexlab{b}})}]{Zammert2014b}
\bibinfo{author}{\bibfnamefont{S.}~\bibnamefont{Zammert}} \bibnamefont{and}
  \bibinfo{author}{\bibfnamefont{B.}~\bibnamefont{Eckhardt}},
  \bibinfo{journal}{J. Fluid Mech.} \textbf{\bibinfo{volume}{761}},
  \bibinfo{pages}{348} (\bibinfo{year}{2014}{\natexlab{b}}).

\bibitem[{\citenamefont{Schneider
  et~al.}(2010{\natexlab{a}})\citenamefont{Schneider, Gibson, and
  Burke}}]{Schneider2010a}
\bibinfo{author}{\bibfnamefont{T.~M.} \bibnamefont{Schneider}},
  \bibinfo{author}{\bibfnamefont{J.~F.} \bibnamefont{Gibson}},
  \bibnamefont{and} \bibinfo{author}{\bibfnamefont{J.}~\bibnamefont{Burke}},
  \bibinfo{journal}{Phys. Rev. Lett.} \textbf{\bibinfo{volume}{104}},
  \bibinfo{pages}{104501} (\bibinfo{year}{2010}{\natexlab{a}}).

\bibitem[{\citenamefont{Schneider
  et~al.}(2010{\natexlab{b}})\citenamefont{Schneider, Marinc, and
  Eckhardt}}]{Schneider2010}
\bibinfo{author}{\bibfnamefont{T.~M.} \bibnamefont{Schneider}},
  \bibinfo{author}{\bibfnamefont{D.}~\bibnamefont{Marinc}}, \bibnamefont{and}
  \bibinfo{author}{\bibfnamefont{B.}~\bibnamefont{Eckhardt}},
  \bibinfo{journal}{J. Fluid Mech.} \textbf{\bibinfo{volume}{646}},
  \bibinfo{pages}{441} (\bibinfo{year}{2010}{\natexlab{b}}).

\bibitem[{\citenamefont{Brand and Gibson}(2014)}]{Brand2014}
\bibinfo{author}{\bibfnamefont{E.}~\bibnamefont{Brand}} \bibnamefont{and}
  \bibinfo{author}{\bibfnamefont{J.~F.} \bibnamefont{Gibson}},
  \bibinfo{journal}{J. Fluid Mech.} \textbf{\bibinfo{volume}{750}},
  \bibinfo{pages}{R1} (\bibinfo{year}{2014}).

\bibitem[{\citenamefont{Chantry et~al.}(2014)\citenamefont{Chantry, Willis, and
  Kerswell}}]{Chantry2013}
\bibinfo{author}{\bibfnamefont{M.}~\bibnamefont{Chantry}},
  \bibinfo{author}{\bibfnamefont{A.~P.} \bibnamefont{Willis}},
  \bibnamefont{and} \bibinfo{author}{\bibfnamefont{R.~R.}
  \bibnamefont{Kerswell}}, \bibinfo{journal}{Phys. Rev. Lett.}
  \textbf{\bibinfo{volume}{112}}, \bibinfo{pages}{164501}
  (\bibinfo{year}{2014}).

\bibitem[{\citenamefont{Gibson}(2012)}]{J.F.Gibson2012}
\bibinfo{author}{\bibfnamefont{J.~F.} \bibnamefont{Gibson}},
  \bibinfo{type}{Tech. Rep.}, \bibinfo{institution}{U. New Hampshire}
  (\bibinfo{year}{2012}), \urlprefix\url{Channelflow.org}.

\bibitem[{\citenamefont{Viswanath}(2007)}]{Viswanath2007}
\bibinfo{author}{\bibfnamefont{D.}~\bibnamefont{Viswanath}},
  \bibinfo{journal}{J. Fluid Mech.} \textbf{\bibinfo{volume}{580}},
  \bibinfo{pages}{339} (\bibinfo{year}{2007}).

\bibitem[{\citenamefont{Guennebaud et~al.}(2010)\citenamefont{Guennebaud,
  Jacon, and Others}}]{eigenweb}
\bibinfo{author}{\bibfnamefont{G.}~\bibnamefont{Guennebaud}},
  \bibinfo{author}{\bibfnamefont{B.}~\bibnamefont{Jacon}}, \bibnamefont{and}
  \bibinfo{author}{\bibnamefont{Others}}, \emph{\bibinfo{title}{{Eigen v3}}}
  (\bibinfo{year}{2010}), \urlprefix\url{http://eigen.tuxfamily.org}.

\bibitem[{\citenamefont{Skufca et~al.}(2006)\citenamefont{Skufca, Yorke, and
  Eckhardt}}]{Skufca2006}
\bibinfo{author}{\bibfnamefont{J.}~\bibnamefont{Skufca}},
  \bibinfo{author}{\bibfnamefont{J.~A.} \bibnamefont{Yorke}}, \bibnamefont{and}
  \bibinfo{author}{\bibfnamefont{B.}~\bibnamefont{Eckhardt}},
  \bibinfo{journal}{Phys. Rev. Lett.} \textbf{\bibinfo{volume}{96}},
  \bibinfo{pages}{174101} (\bibinfo{year}{2006}).

\bibitem[{\citenamefont{Schneider et~al.}(2007)\citenamefont{Schneider,
  Eckhardt, and Yorke}}]{Schneider2007}
\bibinfo{author}{\bibfnamefont{T.~M.} \bibnamefont{Schneider}},
  \bibinfo{author}{\bibfnamefont{B.}~\bibnamefont{Eckhardt}}, \bibnamefont{and}
  \bibinfo{author}{\bibfnamefont{J.}~\bibnamefont{Yorke}},
  \bibinfo{journal}{Phys. Rev. Lett.} \textbf{\bibinfo{volume}{99}},
  \bibinfo{pages}{034502} (\bibinfo{year}{2007}).

\bibitem[{\citenamefont{Melnikov et~al.}(2014)\citenamefont{Melnikov, Kreilos,
  and Eckhardt}}]{Melnikov2014}
\bibinfo{author}{\bibfnamefont{K.}~\bibnamefont{Melnikov}},
  \bibinfo{author}{\bibfnamefont{T.}~\bibnamefont{Kreilos}}, \bibnamefont{and}
  \bibinfo{author}{\bibfnamefont{B.}~\bibnamefont{Eckhardt}},
  \bibinfo{journal}{Phys. Rev. E} \textbf{\bibinfo{volume}{89}},
  \bibinfo{pages}{043008} (\bibinfo{year}{2014}).

\bibitem[{\citenamefont{Nagata}(1997)}]{Nagata1997}
\bibinfo{author}{\bibfnamefont{M.}~\bibnamefont{Nagata}},
  \bibinfo{journal}{Phys. Rev. E} \textbf{\bibinfo{volume}{55}},
  \bibinfo{pages}{2023} (\bibinfo{year}{1997}).

\bibitem[{\citenamefont{Gradshteyn and Ryzhik}(2007)}]{Jeffrey}
\bibinfo{author}{\bibfnamefont{I.~S.} \bibnamefont{Gradshteyn}}
  \bibnamefont{and} \bibinfo{author}{\bibfnamefont{I.~M.}
  \bibnamefont{Ryzhik}}, \emph{\bibinfo{title}{{Table of Integrals, Series, and
  Products}}} (\bibinfo{publisher}{Academic Press}, \bibinfo{year}{2007}),
  \bibinfo{edition}{7th} ed.

\bibitem[{\citenamefont{Fl{\"{u}}gge}(1999)}]{Flugge1999}
\bibinfo{author}{\bibfnamefont{S.}~\bibnamefont{Fl{\"{u}}gge}},
  \emph{\bibinfo{title}{{Practical Quantum Mechanics}}}
  (\bibinfo{publisher}{Springer}, \bibinfo{address}{Berlin/Heidelberg},
  \bibinfo{year}{1999}).

\bibitem[{\citenamefont{Khapko et~al.}(2016)\citenamefont{Khapko, Kreilos,
  Schlatter, Duguet, Eckhardt, and Henningson}}]{Khapko2016a}
\bibinfo{author}{\bibfnamefont{T.}~\bibnamefont{Khapko}},
  \bibinfo{author}{\bibfnamefont{T.}~\bibnamefont{Kreilos}},
  \bibinfo{author}{\bibfnamefont{P.}~\bibnamefont{Schlatter}},
  \bibinfo{author}{\bibfnamefont{Y.}~\bibnamefont{Duguet}},
  \bibinfo{author}{\bibfnamefont{B.}~\bibnamefont{Eckhardt}}, \bibnamefont{and}
  \bibinfo{author}{\bibfnamefont{D.~S.} \bibnamefont{Henningson}}
  \bibinfo{journal}{Phys. Rev. Fluids} \textbf{\bibinfo{volume}{1}},
  \bibinfo{pages}{043602} (\bibinfo{year}{2016}).


\end{thebibliography}

\end{document}